\documentclass[conference]{IEEEtran}
\usepackage{cite}
\usepackage{graphicx}
\usepackage[cmex10]{amsmath}
\usepackage{amssymb}
\usepackage{mathtools}
\usepackage{multirow}
\usepackage{notoccite}
\usepackage{color} 
\usepackage{epsfig}
\usepackage{latexsym}
\usepackage{pstricks}
\usepackage{pdfpages}
\usepackage{epstopdf}
\DeclareGraphicsExtensions{.eps}
\usepackage{algorithmic}
\usepackage{algorithm}
\usepackage{multirow}
\usepackage{graphicx}
\ifCLASSINFOpdf
\else
\fi

\begin{document}
	\title{Successive Decode-and-Forward Relaying with Reconfigurable Intelligent Surfaces}
	
		\author{\IEEEauthorblockN{Zaid Abdullah, Steven Kisseleff, Konstantinos Ntontin, Wallace Alves Martins,\\ Symeon Chatzinotas, and Bj$\ddot{\text{o}}$rn Ottersten \\
			\\ Interdisciplinary Centre for Security, Reliability and Trust (SnT), University of Luxembourg, Luxembourg. \\
			E-mails: \{zaid.abdullah,  steven.kisseleff, kostantinos.ntontin, wallace.alvesmartins,\\ symeon.chatzinotas,  bjorn.ottersten\}@uni.lu} }
	
	\maketitle
	
	\begin{abstract}
	The key advantage of successive relaying (SR) networks is their ability to mimic the full-duplex (FD) operation with half-duplex (HD) relays. However, the main challenge that comes with such schemes is the associated inter-relay interference (IRI). In this work, we propose a reconfigurable intelligent surface (RIS)-enhanced SR network, where one RIS is deployed near each of the two relay nodes to provide spatial suppression of IRI, and to maximize the gain of desired signals. The resultant max-min optimization problem with joint phase-shift design for both RISs is first tackled via the semidefinite programming (SDP) approach. Then, a lower-complexity solution suitable for real-time implementation is proposed based on particle swarm optimization (PSO).  Numerical results demonstrate that even relatively small RISs can provide significant gains in achievable rates of SR networks, and the proposed PSO scheme can achieve a near optimal performance.
	\end{abstract}
	
	
	\IEEEpeerreviewmaketitle

\section{Introduction}
Successive relaying (SR) is a well known technique in wireless communication, where two half-duplex (HD) relays simultaneously assist the transmission between a source and destination nodes utilizing the full bandwidth, and thus imitating full-duplex (FD) operation \cite{fan2007recovering}. The main idea is that at a given time-instant, the source transmits data to one relay, and simultaneously, the other relay forwards the data that it received from the source in the previous transmission time-instant to the destination. Hence, it becomes quite clear that a main challenge in realizing such a scheme is inter-relay interference (IRI). Thus far, authors have adopted different approaches to deal with IRI, such as channel ordering and rate adaptation \cite{chen2014cao}, relay interference cancellation and precoding \cite{wicaksana2009af}, or by neglecting IRI assuming fixed directional relays or large distance between active relays \cite{ikhlef2012mimicking}. 
\par However, the recently proposed reconfigurable intelligent surfaces (RISs) can be a game changer for many wireless applications, including SR as we will demonstrate in this paper. Thanks to their ability in tweaking the wireless environment while maintaining low cost and power requirements, RISs have gained much attention as a strong candidate for future wireless networks \cite{tan2016increasing, rajatheva2020white, chowdhury20206g, kisseleff2020reconfigurable}. In principle, RISs are similar to FD amplify-and-forward relaying, with the former providing passive beamforming without the need to include active power amplifiers or radio-frequency chains, while the latter has the ability to provide active amplification at the cost of increased hardware complexity and power consumption. 
\par Instead of dealing with conventional active relaying and RISs as two separate  technologies, few works have demonstrated the benefits of integrating RISs in active relaying networks. In particular, the works in \cite{abdullah2020hybrid, abdullah2020optimization, wang2021joint, yildirim2021hybrid, obeed2021joint} have shown that RISs can considerably improve the effective rates of conventional relay-based networks. However, in all of these works the authors considered a conventional relay network with a single relay node aided by an RIS, and no SR or IRI was considered in any of these works. In contrast, in this work we investigate the performance of SR networks with the help of two RISs, where we jointly design the passive beamforming weights for the two RISs to provide a spatial (on the air) suppression of the IRI while maximizing the desired signals gains. Our contributions can be summarized as follows:
\begin{itemize}
	\item We propose a new successive decode-and-forward (DF) relaying network with two RISs, and investigate its effective rate performance. The corresponding max-min optimization problem is formulated and solved via a semidefinite programming (SDP) approach. 
	\item In addition, as SDP schemes are known to suffer from a high computational complexity, we propose an efficient evolutionary scheme based on particle swarm optimization (PSO) \cite{kennedy1995particle} to tackle the joint phase-shift design for both RISs. Furthermore, modifications of the PSO are proposed exploiting the structure of the problem to guarantee stability and control the speed and accuracy of convergence.     
	\item Our results demonstrate that RISs can suppress IRI arising from the SR, and a large gain in achievable rates is obtained compared to the cases where only relays or RISs are utilized. Moreover, the proposed PSO scheme is shown to be highly efficient and can achieve a near-optimal performance when compared to the SDP solution.
\end{itemize}It is worth highlighting that our proposed PSO scheme is novel in the sense that it was designed specifically to tackle the RIS passive beamforming with guaranteed stability and controlled convergence behavior. 
\par The rest of this paper is organized as follows: The adopted system model is introduced in Section \ref{section2}. Section \ref{section3} deals with the SDP-based beamforming design, while the PSO-based scheme is introduced and explained in detail in Section \ref{section4}. Numerical results are presented and discussed in Section \ref{section5}. Finally, conclusions are drawn in Section \ref{section6}.
\par \textit{Notations}: Matrices and vectors are denoted by uppercase and lowercase boldface letters, respectively. $(.)^\ast,\ (.)^T,\ \text{and}\  (.)^H$ denote the conjugate, transpose, and Hermitian transpose operators, respectively. In addition, $\boldsymbol a_i$, $\left[\boldsymbol A\right]_{:,j}$, and $\left[\boldsymbol A\right]_{i,j}$ are the $i$th row, $j$th column, and $i$th element of the $j$th column of $\boldsymbol A$, respectively, while $\left[\boldsymbol x\right]_i$ and $x_i$ both represent the $i$th element of $\boldsymbol x$ and used interchangeably as appropriate. $\boldsymbol 0_{N\times M}$ and $\boldsymbol I_N$ are the $N\times M$ all-zero and $N\times N$ identity matrices, respectively. Furthermore, $\text{diag}\{\boldsymbol a\}$ is a diagonal matrix whose diagonal is the elements of $\boldsymbol a$, while $\text{diag}\{\boldsymbol A\}$ is a vector whose elements are the diagonal elements of $\boldsymbol A$. The trace of $\boldsymbol A$ is denoted by $\text{tr}(\boldsymbol A)$, while $\boldsymbol A \succeq 0$ indicates that $\boldsymbol A$ is a positive semidefinite matrix. Finally, $\left|.\right|$ and $\mathbb E\{.\}$ denote the absolute and expectation operators, respectively. 
\section{System model} \label{section2} 
We consider a time-division duplex scenario where a source node ($S$) transmits data to a destination node ($D$). Two HD-DF relays $R_1$ and $R_2$, located between $S$ and $D$, assist the communication by adopting SR to exploit the full bandwidth. Furthermore, two RISs denoted by $I_1$ and $I_2$, located near $R_1$ and $R_2$, respectively, are utilized to enhance the communication quality and suppress the IRI arising from the SR operation. We assume that each of $S$, $D$, $R_1$ and $R_2$ are equipped with a single omni-directional antenna, while $I_1$ and $I_2$ each has $M$ reflective elements.  
\par At any given transmission-time slot, one of the two relays operates as a receiver (Rx) to receive the new block of data from $S$, while the other relay operates as a transmitter (Tx) to forward the decoded message received from $S$ in the previous time instant to $D$; see Fig. \ref{system}.\footnote{Note that during the first and last transmission time slots, only a single relay will be active and no IRI will be present. However, in this work we focus on the case where both relays are active.} Focusing on the case where $R_1$ operates as an Rx, and assuming that signals reflected from each RIS more than once have very low powers and thus can be neglected, we can write the received signal at $R_1$ as:
{\small \begin{align}
	y_{r_1} & = \sqrt{p_s} \left(h_{sr_1}+\boldsymbol{h}_{i_1r_1}^T\boldsymbol \Theta_1\boldsymbol{h}_{si_1}+  \boldsymbol{h}_{i_2r_1}^T\boldsymbol \Theta_2\boldsymbol{h}_{si_2}\right)x \nonumber \\ & + \sqrt{p_{r_2}} \left(h_{r_2r_1}+\boldsymbol{h}_{i_1r_1}^T\boldsymbol \Theta_1\boldsymbol{h}_{r_2i_1}+  \boldsymbol{h}_{i_2r_1}^T\boldsymbol \Theta_2\boldsymbol{h}_{r_2i_2}\right)\tilde{x} + w_{r_1},
\end{align}}where $p_s$ and $p_{r_2}$ are the transmit powers of $S$ and $R_2$, respectively, $\boldsymbol \Theta_1\in \mathbb C^{M\times M}$ and $\boldsymbol \Theta_2\in \mathbb C^{M\times M}$ are the diagonal reflection matrices of $I_1$ and $I_2$, respectively, while $h_{sr_1}\in\mathbb C$, $h_{r_2r_1}\in\mathbb C$, $\boldsymbol h_{i_1r_1} \in \mathbb C^{M\times 1}$, $\boldsymbol h_{si_1} \in \mathbb C^{M\times 1}$, $\boldsymbol h_{i_2r_1} \in \mathbb C^{M\times 1}$, $\boldsymbol h_{si_2} \in \mathbb C^{M\times 1}$, $\boldsymbol h_{r_2i_1} \in \mathbb C^{M\times 1}$,  and $\boldsymbol h_{r_2i_2} \in \mathbb C^{M\times 1}$ denote the channels between $S\rightarrow R_1$, $R_2\rightarrow R_1$, $I_1\rightarrow R_1$, $S\rightarrow I_1$, $I_2\rightarrow R_1$, $S\rightarrow I_2$, $R_2\rightarrow I_1$, and $R_2\rightarrow I_2$, respectively.\footnote{The different narrowband fading channels adopted in this work will be explained in detail in Section \ref{section5}.} Moreover, $x$ and $\tilde{x}$ are the information symbols transmitted from $S$ and $R_2$, respectively, with $\mathbb E\{|x|^2\} = \mathbb E\{|\tilde x|^2\}=1$, and $w_{r_1} \sim \mathcal {CN} (0, \sigma^2)$ is the additive white Gaussian noise (AWGN) at $R_1$.\footnote{Throughout this work, we assume perfect channel state information (CSI) is available similar to \cite{wu2019intelligent} and \cite{huang2019reconfigurable}. Moreover, we assume a centralized processing such that all channels are known to either $S$ or $D$ in order to perform the joint phase-shift design.} \par For more convenience, we express the effective channels between each node ($S$, $R_1$, and $R_2$) and the two RISs in a single vector. Specifically, we define $\small \boldsymbol h_{si} = \begin{bmatrix} \boldsymbol h_{si_1}\\ \boldsymbol h_{si_2} \end{bmatrix} \in\mathbb C^{2M\times 1}$, $\small \boldsymbol h_{ir_1} = \begin{bmatrix} \boldsymbol h_{i_1r_1}\\ \boldsymbol h_{i_2r_1} \end{bmatrix} \in\mathbb C^{2M\times 1}$, $\small \boldsymbol h_{r_2i} = \begin{bmatrix} \boldsymbol h_{r_2i_1}\\ \boldsymbol h_{r_2i_2} \end{bmatrix} \in\mathbb C^{2M\times 1}$, $\small \boldsymbol \theta = \begin{bmatrix} \boldsymbol \theta_1 \\ \boldsymbol \theta_2 \end{bmatrix} \in\mathbb C^{2M\times 1}$, where   $\boldsymbol \theta_j = \text{diag}\{\boldsymbol \Theta_j\} \  (j\in\{1,2\})$. Hence, the signal-to-interference plus noise ratio (SINR) at $R_1$ can be expressed as follows:
\begin{figure}[t]
	\centering
	\hspace*{0.25cm}
	{\includegraphics[width=8cm,height=6.5cm, trim={0cm 0cm 0cm 0cm},clip]{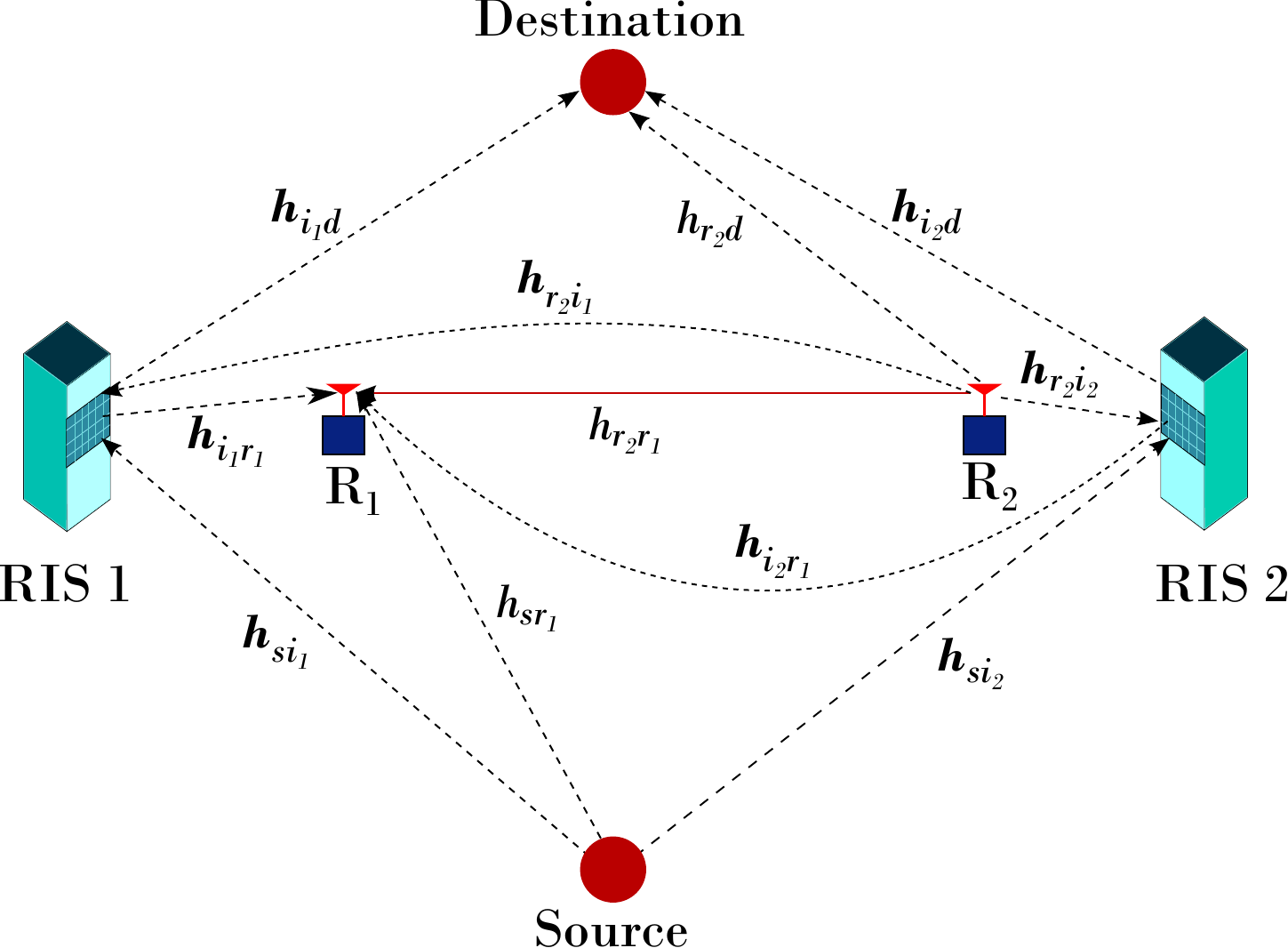}}
	\caption{The proposed RIS-enhanced SR scheme when $R_1$ operates as an Rx while $R_2$ operates as a Tx.}
	\label{system}
\end{figure}
\begin{equation} \label{gamma_r}
	\small 
	\gamma_{r_1} = \frac{p_s \left|h_{sr_1} + \boldsymbol h_{ir_1}^T \boldsymbol \Theta \boldsymbol h_{si}\right|^2}{p_{r_2} \left|h_{r_2r_1}+\boldsymbol h_{ir_1}^T\boldsymbol \Theta \boldsymbol h_{r_2i} \right|^2 + \sigma^2},
\end{equation}where $\boldsymbol \Theta = \text{diag}\{\boldsymbol \theta\} \in \mathbb C^{2M\times 2M}$. On the other hand, we can write the received signal at $D$ (assuming no direct link between $S$ and $D$ exists except through the RISs) as follows:
\begin{equation}
	\small 
	y_d = \sqrt{p_{r_2}} \left( h_{r_2d} + \boldsymbol h_{id}^T \boldsymbol \Theta \boldsymbol h_{r_2i}\right)\tilde{x} + \sqrt{p_{s}} \left( \boldsymbol h_{id}^T \boldsymbol \Theta \boldsymbol h_{si}\right){x}  + w_d, 
\end{equation}where $\small \boldsymbol h_{id} = \begin{bmatrix} \boldsymbol h_{i_1d}\\ \boldsymbol h_{i_2d} \end{bmatrix} \in\mathbb C^{2M\times 1}$ is a vector containing the channel coefficients between $D$ and both RISs, $h_{r_2d}$ is the channel coefficient between $R_2$ and $D$, and $w_d \sim \mathcal{CN} (0, \sigma^2)$ is the AWGN at $D$. Noting that $x$ represents interference to the destination in this case, we can write the SINR at $D$ as:
\begin{equation}\label{gamma_d}
	\small
	\gamma_d = \frac{p_{r_2} \left|h_{r_2d} + \boldsymbol h_{id}^T \boldsymbol \Theta \boldsymbol h_{r_2i}\right|^2}{p_s \left| \boldsymbol h_{id}^T \boldsymbol \Theta \boldsymbol h_{si}\right|^2 + \sigma^2}.
\end{equation}In the subsequent time-instant, $S$ will transmit a new block of data to $R_2$ which will operate as an Rx, whereas $R_1$ will forward its received signal from $S$ in the current time-slot, assuming successful decoding, to $D$. Note that if the distances between $\{S,D\}$ and $\{R_i, I_i\}$ were the same for both $i\in \{1,2\}$, then the system is symmetric, and thus it is sufficient to consider only a single phase to evaluate the achievable rate (in this case $R_1$ is the Rx while $R_2$ is the Tx). Therefore, as we adopt a symmetric system model in this work \cite{zhang2009achievable}, we can write the effective rate at $D$, in bits/s/Hz, as follows:
\begin{equation} \label{rate} 
	\small 
	\mathcal R = \min \left\{\log_2\left(1+\gamma_{r_1}\right), \ \log_2 \left(1+\gamma_d\right)\right\}.
\end{equation}It is clear from (\ref{gamma_r}) and (\ref{gamma_d}) that the achievable rate depends on the reflection matrix $\boldsymbol \Theta$. In the next section, we formulate and solve the joint passive beamforming design problem by utilizing an SDP approach. 
\section{Passive Beamforming Design via Semidefinite Programming} \label{section3}
In this section, we aim to maximize the achievable rate in (\ref{rate}) via an SDP approach. We start by formulating the corresponding optimization problem as follows:
\begin{subequations} \label{OP1}
\small
\begin{align}
	& \hspace{.75cm}\underset{\boldsymbol \Theta}{\text{maximize}} \hspace{.3cm} \min  \left\{\log_2\left(1+\gamma_{r_1}\right), \ \log_2 \left(1+\gamma_d\right)\right\}  \hspace{1cm} (\ref{OP1}) \nonumber \\
	&\hspace{.75cm}\text{subject to} \nonumber \\ 
	& \hspace{.75cm} \big|[\boldsymbol{\Theta}]_{m,m}\big| =1,  \hspace{0.3cm}\forall m\in\mathcal M, \label{6a} 
\end{align}
\end{subequations}where $\mathcal M = \{1, 2, ..., 2M\}$. Due to the interference terms in (\ref{gamma_r}) and (\ref{gamma_d}), and the unit-modulus constraint on $\boldsymbol \Theta$ in (\ref{6a}), problem (\ref{OP1}) is non-convex and cannot be solved in its current form in polynomial time. Therefore, in the following we adopt change of variables and introduce exponential slack variables to overcome the non-convexity of (\ref{OP1}).
\par Let $\small \boldsymbol q_{sr_1} = \begin{bmatrix} \text{diag}\{\boldsymbol h_{ir_1}\} \boldsymbol h_{si}\\ h_{sr_1} \end{bmatrix}$, $\small \boldsymbol q_{r_2r_1} = \begin{bmatrix} \text{diag}\{\boldsymbol h_{ir_1}\} \boldsymbol h_{r_2i}\\ h_{r_2r_1} \end{bmatrix}$, 
$\small \boldsymbol q_{r_2d} = \begin{bmatrix} \text{diag}\{\boldsymbol h_{id}\} \boldsymbol h_{r_2i}\\ h_{r_2d} \end{bmatrix}$, $\small \boldsymbol q_{sd} = \begin{bmatrix} \text{diag}\{\boldsymbol h_{id}\} \boldsymbol h_{si}\\ 0 \end{bmatrix}$, and $\small \boldsymbol v = \begin{bmatrix} \boldsymbol \theta \\ 1 \end{bmatrix}$. Moreover, by defining $\boldsymbol Q_{sr_1} = \boldsymbol q_{sr_1}\boldsymbol q_{sr_1}^H$, $\boldsymbol Q_{r_2r_1} = \boldsymbol q_{r_2r_1}\boldsymbol q_{r_2r_1}^H$, $\boldsymbol Q_{r_2d} = \boldsymbol q_{r_2d}\boldsymbol q_{r_2d}^H$, $\boldsymbol Q_{sd} = \boldsymbol q_{sd}\boldsymbol q_{sd}^H$, and $\boldsymbol V = \boldsymbol v^\ast \boldsymbol v^T$, we can now equivalently represent $\gamma_{r_1}$ and $\gamma_d$ as follows:
\begin{subequations}
	\begin{equation}
		\small 
		\gamma_{r_1} = \frac{p_s \text{tr}\left(\boldsymbol V \boldsymbol Q_{sr_1}\right)}{p_{r_2} \text{tr}\left(\boldsymbol V \boldsymbol Q_{r_2r_1}\right) + \sigma^2},
	\end{equation}
\begin{equation}
	\small 
	\gamma_{d} = \frac{p_{r_2} \text{tr}\left(\boldsymbol V \boldsymbol Q_{r_2d}\right)}{p_s \text{tr}\left(\boldsymbol V \boldsymbol Q_{sd}\right) + \sigma^2}.
\end{equation}
\end{subequations}However, even with relaxing the rank-one constraint on $\boldsymbol V$, formulating the max-min optimization problem with the current formulations of $\gamma_{r_1}$ and $\gamma_{d}$ would still result in a non-convex objective function. Nonetheless, this can now be dealt with by introducing additional slack variables \cite{zhao2015robust}. In particular, let $\boldsymbol s = [s_1\ \  s_2]^T$ and $\boldsymbol u = [u_1\ \ u_2 ]^T$ be exponential slack variables such that
\begin{subequations}
	\begin{equation}
		\small 
		p_s \text{tr}\left(\boldsymbol V \boldsymbol Q_{sr_1}\right) + p_{r_2} \text{tr}\left(\boldsymbol V \boldsymbol Q_{r_2r_1}\right) + \sigma^2 = e^{s_1}, 
	\end{equation}
	\begin{equation}
	\small 
	p_{r_2} \text{tr}\left(\boldsymbol V \boldsymbol Q_{r_2r_1}\right) + \sigma^2 = e^{u_1}, 
\end{equation}
	\begin{equation}
		\small 
		p_{r_2} \text{tr}\left(\boldsymbol V \boldsymbol Q_{r_2d}\right) + p_s \text{tr}\left(\boldsymbol V \boldsymbol Q_{sd}\right) + \sigma^2 = e^{s_2}, 
	\end{equation}
	\begin{equation}
		\small 
		 p_s \text{tr}\left(\boldsymbol V \boldsymbol Q_{sd}\right) + \sigma^2 = e^{u_2}.
	\end{equation}
\end{subequations}Subsequently, and after dropping the rank-one constraint on $\boldsymbol V$ via semidefinite relaxation \cite{luo2010semidefinite}, we can introduce the following optimization problem\footnote{It is worth highlighting that $\frac{1}{\ln(2)}$ is omitted from the objective function without any loss of optimality.}
\begin{subequations} \label{OP2}
	{\small \begin{align}
		& \hspace{.85cm}\underset{ \substack {s_1,\  s_2,\  u_1, \\  u_2, \ \boldsymbol V}}{\text{maximize}} \hspace{.5cm} \underset{i\in\{1,2\}}{\min}  \hspace{.5cm} \left\{s_i - u_i\right\}  \hspace{2.5cm} (\ref{OP2}) \nonumber \\
		&\hspace{.85cm}\text{subject to} \nonumber \\ 
		&\hspace{.85cm} 	p_s \text{tr}\left(\boldsymbol V \boldsymbol Q_{sr_1}\right) + p_{r_2} \text{tr}\left(\boldsymbol V \boldsymbol Q_{r_2r_1}\right) + \sigma^2 \ge e^{s_1}, \label {9a}\\
		&\hspace{.85cm} p_{r_2} \text{tr}\left(\boldsymbol V \boldsymbol Q_{r_2r_1}\right) + \sigma^2 \le e^{u_1}, \label {9b} \\
		&\hspace{.85cm}  p_{r_2} \text{tr}\left(\boldsymbol V \boldsymbol Q_{r_2d}\right) + p_s \text{tr}\left(\boldsymbol V \boldsymbol Q_{sd}\right) + \sigma^2 \ge e^{s_2}, \label {9c} \\ 
		&\hspace{.85cm} p_s \text{tr}\left(\boldsymbol V \boldsymbol Q_{sd}\right) + \sigma^2 \le e^{u_2}, \label {9d} \\ 
		&\hspace{.8cm} \left[\boldsymbol V \right]_{m,m} =1,  \ \ m \in \{1, 2, ...., 2M+1\}, \label {9e} \\
		&\hspace{.85cm} \boldsymbol V \succeq 0. \label {9f}
	\end{align}}\end{subequations}The only non-convex constraints at this stage are (\ref{9b}) and (\ref{9d}). To overcome this drawback, a first-order Taylor approximation is applied to convert these constraints into convex ones. In particular, we define $\bar {\boldsymbol u} = [\bar u_1 \ \ \bar u_2]^T$, where $\bar u_1$ and $\bar u_2$ are the points around which the linearization is made such that $e^{u_i} \approx  e^{\bar u_i}(u_i - \bar u_i + 1)$. Hence, we can recast problem (\ref{OP2}) as
\begin{subequations} \label{OP3}
	{\small \begin{align}
			& \hspace{.85cm}\underset{ \substack {s_1,\  s_2,\  u_1, \\  u_2, \ \boldsymbol V}}{\text{maximize}} \hspace{.5cm} \underset{i\in\{1,2\}}{\min}  \hspace{.5cm} \left\{s_i - u_i\right\}  \hspace{2.5cm} (\ref{OP3}) \nonumber \\
			&\hspace{.85cm}\text{subject to} \nonumber \\ 
			&\hspace{.85cm} p_{r_2} \text{tr}\left(\boldsymbol V \boldsymbol Q_{r_2r_1}\right) + \sigma^2 \le e^{\bar u_1}(u_1 - \bar u_1 + 1), \\
			&\hspace{.85cm} p_s \text{tr}\left(\boldsymbol V \boldsymbol Q_{sd}\right) + \sigma^2 \le e^{\bar u_2}(u_2 - \bar u_2 + 1), \\ 
			&\hspace{.85cm}\text{and constraints (\ref{9a}), (\ref{9c}), (\ref{9e}), and (\ref{9f})}.
\end{align}}\end{subequations}It follows that all constraints and objective function at this stage are convex, and problem (\ref{OP3}) can be solved iteratively using software tools such as CVX. However, we firstly initialize $\bar u_1$ and $\bar u_2$ as follows
\begin{subequations}
	\begin{equation}
		\small 
		\bar u^{(0)}_1 = \ln \left(  p_{r_2} \text{tr}\left(\tilde{\boldsymbol V} \boldsymbol Q_{r_2r_1}\right) + \sigma^2 \right),
	\end{equation}
	\begin{equation}
		\small 
		\bar u^{(0)}_2 = \ln \left( p_s \text{tr}\left(\tilde {\boldsymbol V} \boldsymbol Q_{sd}\right) + \sigma^2\right),
	\end{equation}
\end{subequations}where $\tilde{\boldsymbol V}$ is any feasible (i.e. random) solution. Once the algorithm has converged, there is no guarantee that the rank of $\boldsymbol V^\star$ is equal to $1$. In such case, the value of the objective function corresponding to $\boldsymbol V^\star$ represents an upper-bound. Nonetheless, one can apply eigenvalue decomposition (EVD) with Gaussian randomization to approximate a rank-$1$ near optimal solution \cite{wu2018intelligent}. The details are omitted here for brevity, however, it is worth pointing out that such approximation guarantees a minimum accuracy of $\pi/4$ of the optimal objective value \cite{zhang2006complex}. The steps of the proposed scheme are given in Algorithm \ref{Algorithm1}. 
\begin{algorithm}[t]
	\caption{Proposed SDP algorithm for phase-shifts design}
	\label{Algorithm1}
	\begin{algorithmic}[1]
		\STATE \textbf{input} $\mathbf Q_{sr_1}$, $\mathbf Q_{r_2r_1}$, $\mathbf Q_{r_2d}$, $\mathbf Q_{sd}$, $p_s$, $p_{r_2}$, $\sigma^2$, $\bar  u^{(0)}_1$, $\bar  u^{(0)}_2$, $\epsilon = 10^{-3}$ (accuracy), $k=1$ (iteration index),
		\STATE \textbf{define} $\small {\text{err}^{(k)}} \overset {\Delta}{=} \sum_{i=1}^{2}\big|u^{(k)}_i - \bar{u}^{(k-1)}_i\big|$
		\STATE \textbf{repeat}
		\STATE \ \ \ \ \textbf{solve} $(\ref{OP3})$ and then \textbf{evaluate} $\text{err}^{(k)}$,
		\STATE \ \ \ \ \textbf{if} $\text{err}^{(k)} <\epsilon$
		\STATE \ \ \ \ \ \ \ \ \textbf{break}; 
		\STATE \ \ \ \ \textbf{else} update $\bar{u}^{(k)}:= u^{(k)}$, and then \textbf{increment} $k := k + 1$,
		\STATE \textbf{until} convergence
	\end{algorithmic}
\end{algorithm} 
\section{Particle Swarm Optimization-based Beamforming Design} \label{section4}
The main drawback for the SDP approach is that it suffers from a complexity order of $\mathcal O\left((2M+1)^{3.5}\right)$ \cite{cui2019secure}, and hence, it becomes unsuitable when dealing with large optimization problems. Therefore, in this section we propose a lower-complexity evolutionary method to tackle the phase-shift design, which is based on PSO \cite{kennedy1995particle}. 
\par We start by generating an initial population of $N$ particles $\boldsymbol F = \big[\boldsymbol f_1, \ \boldsymbol f_2, \ ...\ , \ \boldsymbol f_N\big]^T \in \mathbb R^{N\times 2M}$ with entries (or phase-shifts) drawn randomly in the interval $[-\pi, \pi]$. We evaluate the fitness of the $n$th particle ($\lambda_n$) as follows:
\begin{equation}
	\small
	\lambda_n = \min \Big\{\gamma_{r_1} \left(\boldsymbol f_n\right), \gamma_d \left(\boldsymbol f_n\right) \Big\}, \ \ \ \forall n \in \{1, ..., N\},
\end{equation}where $\gamma_{r_1} \left(\boldsymbol f_n\right)$ and $\gamma_d \left(\boldsymbol f_n\right)$ are the SINR values at $R_1$ and $D$ given in (\ref{gamma_r}) and (\ref{gamma_d}), respectively, such that
{\small\begin{align} 
	\gamma_i(\boldsymbol f_n) = \left\{\gamma_i \middle| \boldsymbol \Theta = \text{diag}\big\{e^{\jmath \boldsymbol f_n}\big\} \right\}, i\in\{r_1, d\}.
\end{align}}Then, we find the \textit{local best} and \textit{global best} solutions. Specifically, the \textit{global best} solution for the $m$th element of each particle is the $m$th entry of the particle that has the highest fitness value in the population. Thus, the vector of global best solutions is in fact the particle that has the highest fitness in the population, which we denote as $\boldsymbol f_{\text{max}}$. \par In contrast, the \textit{local best} solution for the $m$th element of $\boldsymbol f_n$, denoted by $\left[\boldsymbol L\right]_{n,m}$ ($\boldsymbol L \in \mathbb R^{N\times 2M}$), is the $m$th entry of the particle with the highest fitness value among only the neighbouring particles of $\boldsymbol f_n$, which are $\boldsymbol f_{n+1}$ and $\boldsymbol f_{n-1}$.\footnote{We adopt a ring topology, thus, the nighbouring particles for $\boldsymbol f_1$ are $\boldsymbol f_2$ and $\boldsymbol f_N$, while the neighbours of $\boldsymbol f_N$ are $\boldsymbol f_1$ and $\boldsymbol f_{N-1}$.}
\par Furthermore, the velocities that control the direction of each particle, which we initialize as $\boldsymbol X = \boldsymbol 0_{N\times 2M}$, are updated at each iteration as follows \cite{kennedy1995particle}
{\small \begin{align} \label{vel}
	 \left[\boldsymbol X^{(t+1)}\right]_{n,m} & =  \left[\boldsymbol X^{(t)}\right]_{n,m} + w_1r_1 \Big(\left[\boldsymbol L\right]_{n,m}-\left[\boldsymbol F^{(t)}\right]_{n,m}\Big) \nonumber \\ & + w_2r_2 \Big(\left[\boldsymbol f_{\text{max}}\right]_m - \left[\boldsymbol F^{(t)}\right]_{n,m}\Big),  n\in \mathcal N, m\in \mathcal M.
\end{align}}where $w_1$ and $w_2$ are learning factors, $r_1$ and $r_2$ are random numbers drawn from a uniform distribution with values between $0$ and $1$, and $\mathcal N =\{1, ..., N\}$. To ensure a stable convergence,\footnote{Note that performing the normalization in (\ref{norm}) is crucial, and without it the values of $\boldsymbol X$ can become extremely large (or small) after many iterations, preventing the algorithm from finding any good solution.} we normalize each column of $\boldsymbol X^{(t+1)}$ as
\begin{equation}\label{norm}
	\small 
	\left[\boldsymbol X^{(t+1)}\right]_{:,m} := \mu \times \frac{ \left[\boldsymbol X^{(t+1)}\right]_{:,m}} {\max \left\{\left|\left[\boldsymbol X^{(t+1)}\right]_{:,m}\right|\right\}},
\end{equation} where $\mu \in (0, \pi]$ is an introduced (auxiliary) parameter. Then, the phase-shifts are updated as
\begin{equation} \label{update} 
	\small
	\boldsymbol F^{(t+1)} = \boldsymbol F^{(t)} + \boldsymbol X^{(t+1)}.
\end{equation} It is worth to highlight that the value of $\mu$ in (\ref{norm}) controls the speed and accuracy of convergence, and it represents the highest possible phase-shift difference per reflecting element between any two successive iterations. Finally, we re-adjust the entries of $\boldsymbol F^{(t+1)}$ that are either greater than $\pi$ or less than $-\pi$. Due to the circular symmetry of phase-shifts, the re-adjustment can be carried out as follows
\begin{equation} \label{adjust}
	\small
	\left[ \boldsymbol F^{(t+1)} \right]_{n,m} := \begin{cases}
		\left[ \boldsymbol F^{(t+1)} \right]_{n,m} + 2\pi, \ \ \ \ \text{if} \left[ \boldsymbol F^{(t+1)} \right]_{n,m} < -\pi \\ \left[ \boldsymbol F^{(t+1)} \right]_{n,m} - 2\pi, \ \ \ \ \text{if} \left[ \boldsymbol F^{(t+1)} \right]_{n,m} > \pi.
	\end{cases}
\end{equation}As shown in Algorithm \ref{Algorithm2}, the same procedure (i.e. fitness evaluation, obtaining \textit{local best and global best} solutions, update of velocities followed by normalization, update of phase-shifts of the population, and re-adjustment) will be repeated in subsequent iterations until a maximum number of iterations is reached, and the particle that achieves the highest fitness value in all $t\in\{0, 1, ..., T-1\}$ iterations will be selected as the final solution.
\par When it comes to the per-iteration complexity of the proposed PSO scheme, we can observe that the (fitness evaluation, update of velocity, normalization, update of population, and adjustment), each has a complexity order of $\mathcal O (2MN)$. Thus, and compared to the SDP approach, the PSO becomes particularly attractive when dealing with a large number of reflecting elements, as its complexity scales only linearly with the size of optimization problem.
\begin{algorithm}[t]
	\caption{Proposed PSO algorithm for phase-shifts design}
	\label{Algorithm2}
	\begin{algorithmic}[1]
		\STATE \textbf{input} $\boldsymbol F$ (initial population), $\boldsymbol X$ (initial velocities), $t = 0$, $T$ (maximum number of iterations), and $\mu$,
		\STATE \textbf{while} $t  \le T$
		\STATE \ \ \ \ \textbf{evaluate} the fitness of each particle in $\boldsymbol F$,
		\STATE \  \ \ \ \textbf{find} \textit{global best} and \textit{local best} solutions,
		\STATE \ \ \ \ \textbf{update} the velocities according to (\ref{vel}),
		\STATE \ \ \ \ \textbf{normalize} the new velocities according to (\ref{norm}),
		\STATE \ \ \ \ \textbf{update} the population according to (\ref{update}),
		\STATE \ \ \ \ \textbf{adjust} the new population according to (\ref{adjust}),
		\STATE \ \ \ \ \textbf{increment} $t := t + 1$,
		\STATE \textbf{end while}
		\STATE \textbf{output}: particle with highest fitness value.
	\end{algorithmic}
\end{algorithm} 
\section{Results and Discussion} \label{section5}
We start by introducing the different narrowband channels and parameters involved in this paper. All channels from and to each RIS are assumed to follow Rician distribution with both line-of-sight (LoS) and non-LoS (NLoS) links, such that $\boldsymbol h_j = \sqrt{\frac{k_r}{k_r+1}} \boldsymbol h^{\text{los}}_j + \sqrt{\frac{1}{1+k_r}} \boldsymbol h^{\text{nlos}}_j$, where $k_r$ is the Rician $K$ factor, $\boldsymbol h^{\text{los}}_j \in\mathbb C^{M\times 1}$ is the deterministic LoS channel vector for the $j$th link, with each element having an absolute value of $1/d_j^{{\alpha_{\text{los}}}/2}$ and a random phase between $0$ and $2\pi$, $\boldsymbol h^{\text{nlos}}_j \sim \mathcal {CN}(\boldsymbol 0_{M\times 1}, \boldsymbol I_M d_j^{-{\alpha}_{\text{nlos}}})$ is the Rayleigh distributed NLoS channel vector, where $d_j$ is the distance between the nodes associated with the $j$th link, $j \in \{si_1, i_1r_1, r_2i_1, i_1d, si_2, i_2r_1, r_2i_2, i_2d\}$, while $\alpha_{\text{los}}$ and $\alpha_{\text{nlos}}$ represent the path-loss exponents for LoS and NLoS links, respectively. In contrast, $h_i \sim \mathcal {CN} (0, d_i^{-\alpha_{\text{nlos}}})$ represents the channel coefficient with Rayleigh distribution for links that do not involve any of the RISs, i.e. $i\in\{sr_1, r_2d\}$. However, for the IRI link, $h_{r_2r_1}$, we consider Rician channel to demonstrate the efficiency of RISs in suppressing the interference even when a strong LoS channel exists. 
\par
\begin{figure}[t]
\hspace{1.2cm}
	{\includegraphics[width=7.8cm,height=5.6cm, trim={-1cm -1cm -1cm -1cm},clip]{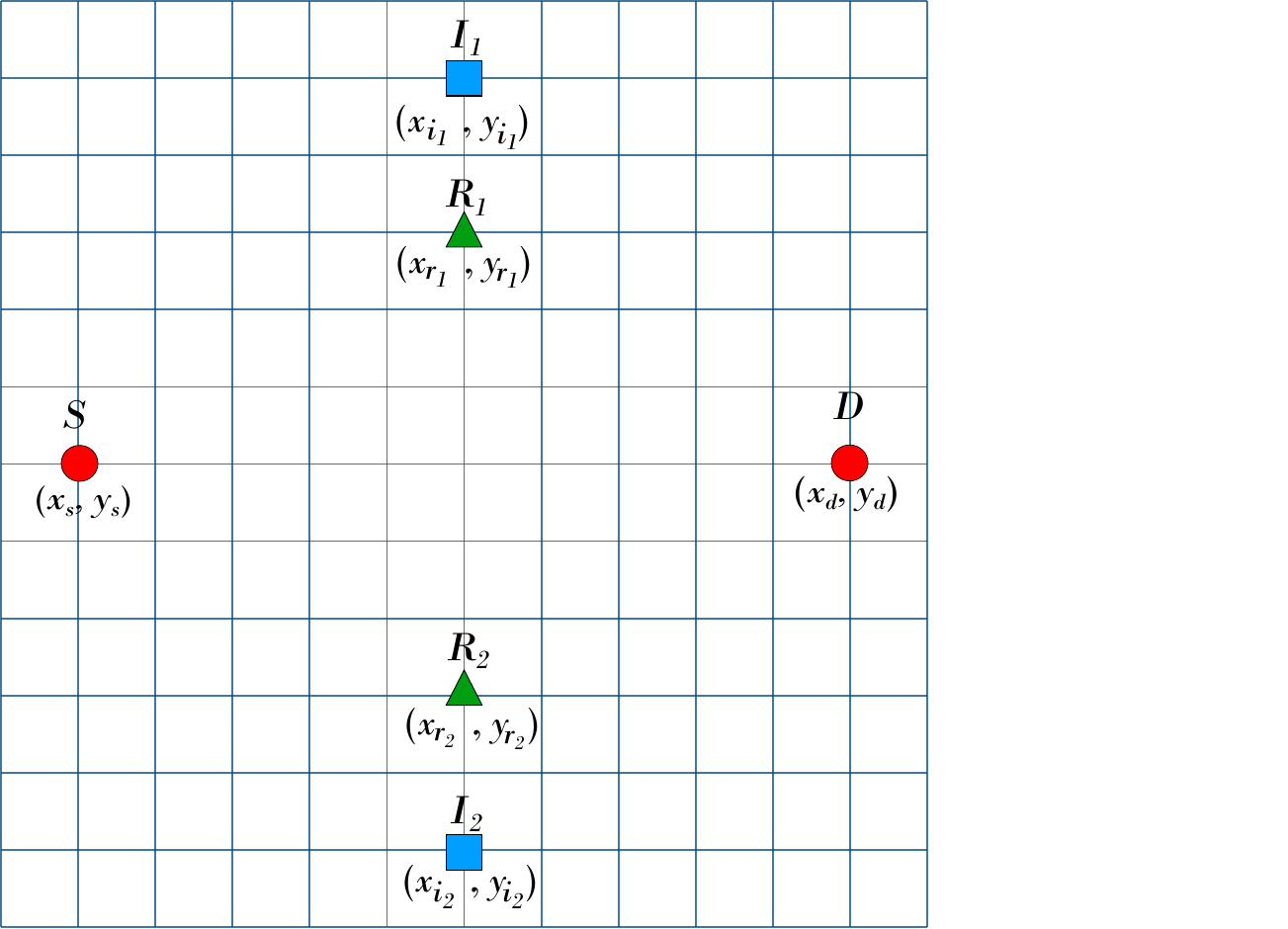}}
	\caption{A top-view of the adopted simulation setup.}
	\label{Fig4}
\end{figure}

 The locations of different nodes are set as: $(x_s, y_s) = (0,0)$, $(x_d, y_d) = (100, 0)$, $(x_{r_1}, y_{r_1}) = (50, 25)$, $(x_{r_2}, y_{r_2}) = (50, -25)$, $(x_{i_1}, y_{i_1}) = (50, 30)$, and $(x_{i_2}, y_{i_2}) = (50, -30)$, all in meter units, see Fig. \ref{Fig4}. In addition, we set $w_1 = w_2 = 2$, $\sigma^2 = 1$, $k_{r} = 5$ dB, $ \alpha_{\text{los}} = 2.3$, and $\alpha_{\text{nlos}} = 3.5$. Equal power allocation was utilized and thus $p_s = p_{r_2} = \frac{1}{2}p$, and we define the transmit signal-to-noise ratio (SNR) as $p/\sigma^2$.
\par Our simulations compare between the following schemes:
\begin{itemize}
	\item \textbf{RISs-assisted SR (upper-bound)}: This refers to the results obtained by SDP without performing EVD with Gaussian randomization, hence, the term upper-bound. Rician fading is assumed for the IRI channel (i.e. $h_{r_2r_1}$).
	\item \textbf{SR without RISs}: This case demonstrates the results for SR utilizing only the two relays without the RISs. Rayleigh fading is adopted for all different channels including the IRI link. 
	\item \textbf{RISs only}: In this case, we only have two RISs (therefore no SR) and we optimize their phase-shifts to reflect the signal from $S$ toward $D$, where the former transmits with a power budget of $p$ Watts.
	\item \textbf{RISs-assisted SR (PSO)}: This case shows the performance of our proposed PSO scheme, assuming Rician-fading for the IRI link.
\end{itemize}
	\begin{figure}[t]
	\centering
	{\includegraphics[width=6.5cm,height=6.5cm, trim={5.3cm 8.2cm 5.3cm 8.2cm},clip]{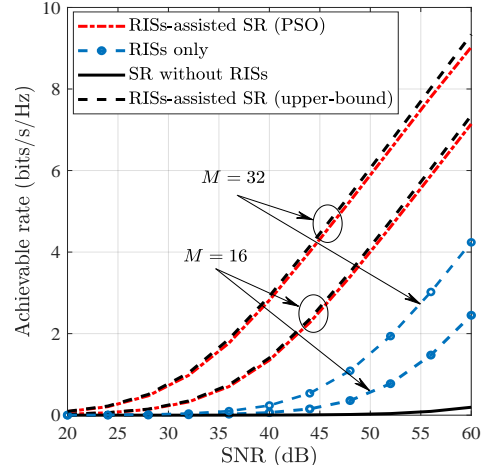}}
	\caption{Achievable rate vs transmit SNR for different schemes when $N = 100$, $T = 200$, and $\mu = \pi/8$.}
	\label{Fig2}
\end{figure}
	\begin{figure}[t]
	\centering
{\includegraphics[width=6.5cm,height=6.5cm, trim={5.3cm 8.2cm 5.3cm 8.2cm},clip]{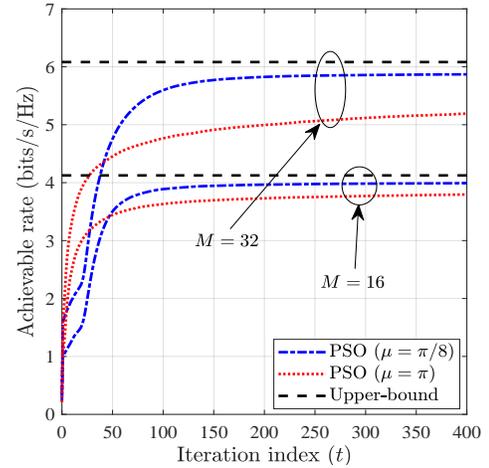}}
	\caption{Convergence behavior of the proposed PSO when $N = 50$, and transmit $\mathrm {SNR} = 50$ dB.}
	\label{Fig3}
\end{figure}
\par In Fig. \ref{Fig2}, we depict the achievable rate vs. SNR curves. The RISs-assisted SR notably outperforms both cases of RISs-only scheme and the SR case without RISs. In particular, to achieve an effective rate of $4$ bits/s/Hz, a gain of $15$ dB in transmit SNR can be obtained for the proposed RISs-assisted SR over the RISs-only approach, given that the number of reflecting elements per RIS is $32$. In contrast, the SR case without RISs shows a very poor performance, which highlights the sever impact of the IRI. Moreover, one can also observe from Fig. \ref{Fig2} that there is only a negligible performance gap between the low-complexity PSO scheme and the more complex SDP approach.
\par In Fig. \ref{Fig3}, we show the convergence behavior of the proposed PSO scheme for different number of reflecting elements and different values of the auxiliary parameter $\mu$. Regardless of the number of reflecting elements, setting $\mu$ to a small value (in this case $\pi/8$) is preferable as it allows the proposed algorithm to converge toward better solutions that are close to the optimal points. In contrast, larger values of $\mu$ enable faster exploration of new and more diverse populations, and thus they are preferable when the number of optimization iterations is small. However, they are more likely to fail in finding near optimal solutions due to the rapid velocity changes of candidate particles, especially when the number of reflecting elements is relatively large.
\section{Conclusion} \label{section6}
In this work, a new RIS-aided relay network architecture was proposed with SR, where two RISs were deployed near the two HD-DF relays to provide a spatial suppression of IRI while maximizing the gain of desired signals. An upper bound solution for the achievable rate was obtained via the SDP technique without performing Gaussian randomization and EVD. Subsequently, a lower-complexity beamforming design based on PSO with controlled convergence speed was proposed, and a new normalization step was introduced which guaranteed the stability of the designed scheme. Numerical results demonstrated that RISs can provide large improvements on the rate performance of SR networks, even when a strong line-of-sight link for the IRI existed. Finally, the proposed PSO scheme showed its capability in obtaining near optimal solutions.   
\section*{Acknowledgment}
This work was supported by the Luxembourg National Research Fund (FNR) under the CORE project RISOTTI.

\bibliographystyle{IEEEtran}
\bibliography{RIS_SR}

\end{document}